\documentclass[aip,apl,twocolumn,reprint,superscriptaddress]{revtex4-1}
\usepackage{graphicx}
\usepackage{color}
\usepackage{url}

\begin{document}
\title{Weak electron-phonon coupling in the early alkali atomic wires}

\author{Nicholas A. Lanzillo$^*$}
\affiliation{Department of Physics, Applied Physics, and Astronomy, Rensselaer Polytechnic Institute, 110 8$^{th}$ Street, Troy, NY 12180, USA}

\author{Saroj K. Nayak}
\affiliation{Department of Physics, Applied Physics, and Astronomy, Rensselaer Polytechnic Institute, 110 8$^{th}$ Street, Troy, NY 12180, USA}
\affiliation{School of Basic Sciences, Indian Institute of Technology, Bhubaneswar, India 751007}

\date{\today}

\begin{abstract}
The structural, electronic and vibrational properties of atomic wires composed of the early alkali metals lithium and sodium are studied using density functional perturbation theory. The s-like electronic states near the Fermi level couple very weakly to longitudinal acoustic phonons and not at all to the transverse acoustic phonons, which results in a weak overall electron-phonon coupling. The results are compared to earlier studies on the electron-phonon coupling in metallic atomic wires and reinforces the idea that s-like states at the Fermi level give rise to weak electron-phonon coupling in one-dimension, in contrast with materials containing d-like states at the Fermi level which have correspondingly larger electron-phonon coupling due to interactions with transverse phonons. 

*Correspondence to lanzin@rpi.edu. 
\end{abstract} 

\maketitle

\section{Introduction}
The interplay between electrons and phonons gives rise to many interesting and important physical properties, including transport phenomena and superconductivity\cite{Ashcroft1976solid}. Some of the earliest attempts at characterizing the electron-phonon interaction from a first-principles perspective using a combination of density functional theory and linear response theory have resulted in remarkably accurate descriptions of phonon frequencies and electron-phonon coupling constants for many common metals\cite{savrasov1994linear,savrasov1996electron,bauer1998electron}. The effects of reduced dimensionality and extreme quantum confinement give rise to distinct electron-phonon coupling relative to the bulk. 

There has been extensive work, both theoretical\cite{Frederiksen2004inelastic,Picaud2003phonons,delaVega2006universal,Agrait2002onset} and experimental\cite{Agrait2002electron,bohler2009point}, on the electron-phonon interaction in short, finite-length, single-atom thick atomic wires. However, there has been comparatively little work done exploring the electron-phonon interaction using Eliashberg Theory in these types of systems. Up to this point, Eliashberg Theory has been used to comparatively study the electron-phonon interactions in Al and Pb\cite{verstraete2006phonon}, Na\cite{sen2006peierls} and Al, Cu, Ag and Au\cite{simbeck2012aluminum}. In this paper, we explore the effects of the electron-phonon interaction in the early alkali metals, Li and Na, which have simple, nearly-spherical Fermi surfaces to paint a more complete picture of electron-phonon coupling in single-atom thick, one-dimensional metallic wires. 

Previous results\cite{verstraete2006phonon,simbeck2012aluminum,sen2006peierls} indicate that there exists a correlation between the electronic character at the Fermi level (i.e. s-like, p-like or d-like) and the type of phonon (longitudinal or transverse) that the electron will couple to. In Al and Ag atomic wires, it is found that the s-like and p-like electrons at the Fermi level couple to longitudinal acoustic phonons and result in an overall weak coupling, while the d-like electrons in Cu, Au and Pb atomic wires couple to imaginary frequency transverse modes, giving rise to a large electron-phonon coupling. In this paper, we extend this understanding to include the electron-phonon coupling in the early alkali metals Li and Na, which possess simple, nearly-spherical Fermi surfaces. 

\section{Theory}
Electronic wavefunctions and energies are calculated via density functional theory\cite{kohn1964inhomogeneous,kohn1965self}, while the phonon dispersion relations and electron-phonon interactions are calculated via density functional perturbation theory\cite{gonze1997dynamical}. The central ingredient in calculating the electron-phonon coupling is the interaction matrix element, $g_{k,k'}$:

\begin{equation}
g_{kk'}=\sqrt{\frac{\hbar}{2NM\omega_q}}\vec{u_q}<k'|\delta V_{SCF}|k>
\end{equation}
where $k$ and $k'$ represent the electronic wavefunctions before and after a phonon collision event, $M$ is the ionic mass, $N$ is the number of phonon modes, $\vec{u_q}$ is the phonon eigenvector and $\delta V_{SCF}$ is the self-consistent change in the potential energy in the presence of the phonon-distorted lattice geometry. The Eliashberg Spectral Function ($\alpha^2F(\omega)$) is then defined as an integral over the matrix elements squared: 

\begin{equation}
\alpha^2F(\omega)=g(\epsilon_F)\sum_{k,k'}|g_{kk'}|^2\delta(\omega-\omega_q)
\end{equation}
where $g(\epsilon_F)$ is the electronic density of states at the Fermi level. This function can be thought of a phonon density of states weighted according to scattering interactions with electrons near the Fermi level. In essence, we keep only the phonons from the density of states that participate in scattering events with electrons. The electron-phonon coupling constant ($\lambda$) is defined as a weighted integral over the Eliashberg Spectral Function:

\begin{equation}
\lambda = 2 \int \frac{\alpha^2F(\omega)}{\omega}
\end{equation}
The factor of $1/\omega$ has the effect of filtering out the contributions from high-frequency phonons to the overall electron-phonon coupling, instead placing most of the weight on the lower-frequencies modes. This makes intuitive sense, since the electronic energy levels before and after a phonon event are related via the phonon frequency ($\omega$):

\begin{equation}
\epsilon_f = \epsilon_i \pm \hbar \omega
\end{equation}
where the + is for phonon absorption and the - is for phonon emission. In general, the change in energy between the initial and final electronic states are small since a typical phonon energy is on the order of meV. Thus, low-frequency phonons are expected to play a greater role in scattering events with electrons than high-frequency phonons.  

\section{Computational Methods} 
Calculations were performed using the ABINIT\cite{gonze2002first,gonze2005brief,gonze2009abinit} software package, which is an implementation of density functional theory\cite{kohn1964inhomogeneous,kohn1965self} utilizing pseudopotentials and a plane-wave basis set. Lithium and sodium atoms were treated using Troullier-Martins norm-conserving pseudopotentials\cite{troullier1991efficient} with plane wave cutoff energies of 80.0 Hartree for Li and 20.0 Hartree for Na. We considered single-atom unit cells with 15.0 Bohr of vacuum in the lateral directions, which is large enough to ensure that the interaction between neighboring supercells in negligible. The interatomic separation was optimized in each case until the forces on the atoms were less than 0.05 eV/Angstrom. We chose k-point sampling of $1\times 1\times64$ for electronic structure calculations and a finer grid of $1 \times 1 \times 128$ of phonon calculations. Convergence with respect to the number of k-points was carefully checked in each case. 

\section{Results and Discussion} 
The starting point for our study was to optimize the interatomic separation for each atomic wire. This was done via two methods; first we minimized the forces on the ions until they were converged below 0.05 eV/Angstrom, and then we calculated potential energy curves within a window of 1-2 atomic units surrounding the minimum energy value to get an idea of the underlying energy landscape. The potential energy curves for our wires are shown in Figure 1. 

\begin{figure}[h!]
\centering
\begin{center}
\includegraphics[scale=0.35,clip,trim = 0mm 40mm 0mm 40mm]{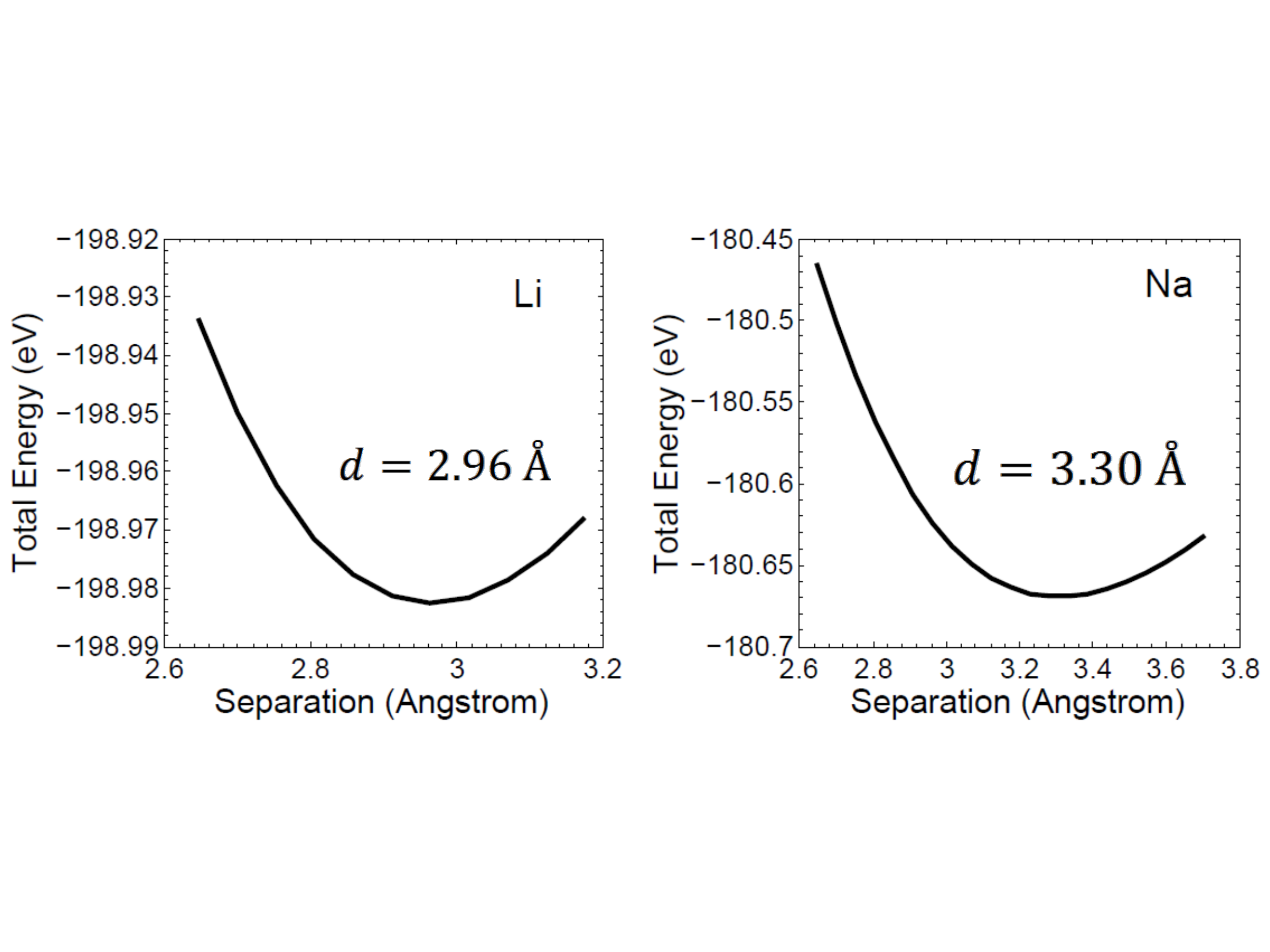}
\end{center}
\caption {The potential energy curves for infinitely-long, linear atomic chains composed of lithium and sodium atoms.}
\end{figure}

Each wire shows a well-defined potential energy minimum for the linear configuration in the neighborhood of 3 Angstrom. While we note that the potential energy minimum for the linear configuration is likely metastable state, similar to the cases of Al, Pb and Na\cite{verstraete2006phonon,sen2006peierls}, searching for other stable geometries like zig-zag configurations is beyond the scope of the current work since we only seek to elucidate the effects of electron-phonon coupling in linear wires. The minimum-energy separation for the lithium wire is 2.96 Angstrom, while for the sodium wire it is 3.30 Angstrom. As expected, the larger sodium atom prefers a larger interatomic separation relative to the smaller lithium atom. 

The electronic energy dispersion is calculated for each wire in order to verify that the wire retains its metallicity when confined to a single spatial dimension and also to investigate the character of energy bands near the Fermi level. The electronic structure and angular momentum resolved density of states for each atomic wire are shown in Figure 2. 

\begin{figure}[h!]
\centering
\begin{center}
\includegraphics[scale=0.35]{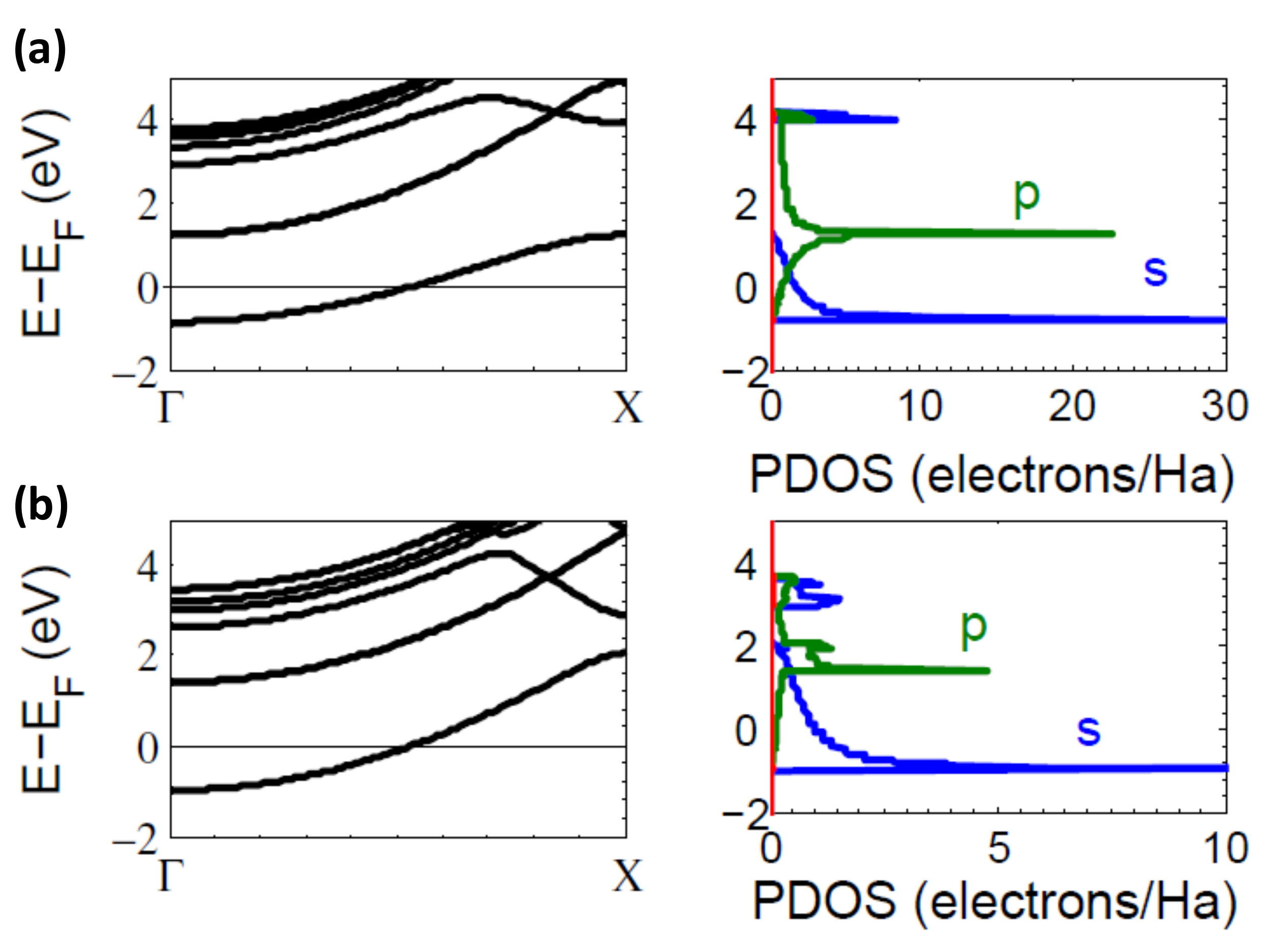}
\end{center}
\caption {The electronic band structure and angular momentum resolved partial density of states for (a) lithium and (b) sodium atomic wires.}
\end{figure}

We see that each wire has a single band crossing the Fermi level and that the character of each band is predominantly s-like at the Fermi level, although the p-like orbitals make a small contribution in the case of lithium. The d-like contributions are not identified because they are unoccupied and lie far enough away from the Fermi level that they don't contribute to the net electron-phonon coupling constant. In both cases, the electron character shifts from s-like below the Fermi level (at the Gamma point) to p-like above the Fermi level (at the X point). 

Next, we turn to the phonons. The phonon band structure, Eliashberg Spectral Function and electron-phonon coupling constant are shown in Figure 3. 

\begin{figure}[h!]
\centering
\begin{center}
\includegraphics[scale=0.35]{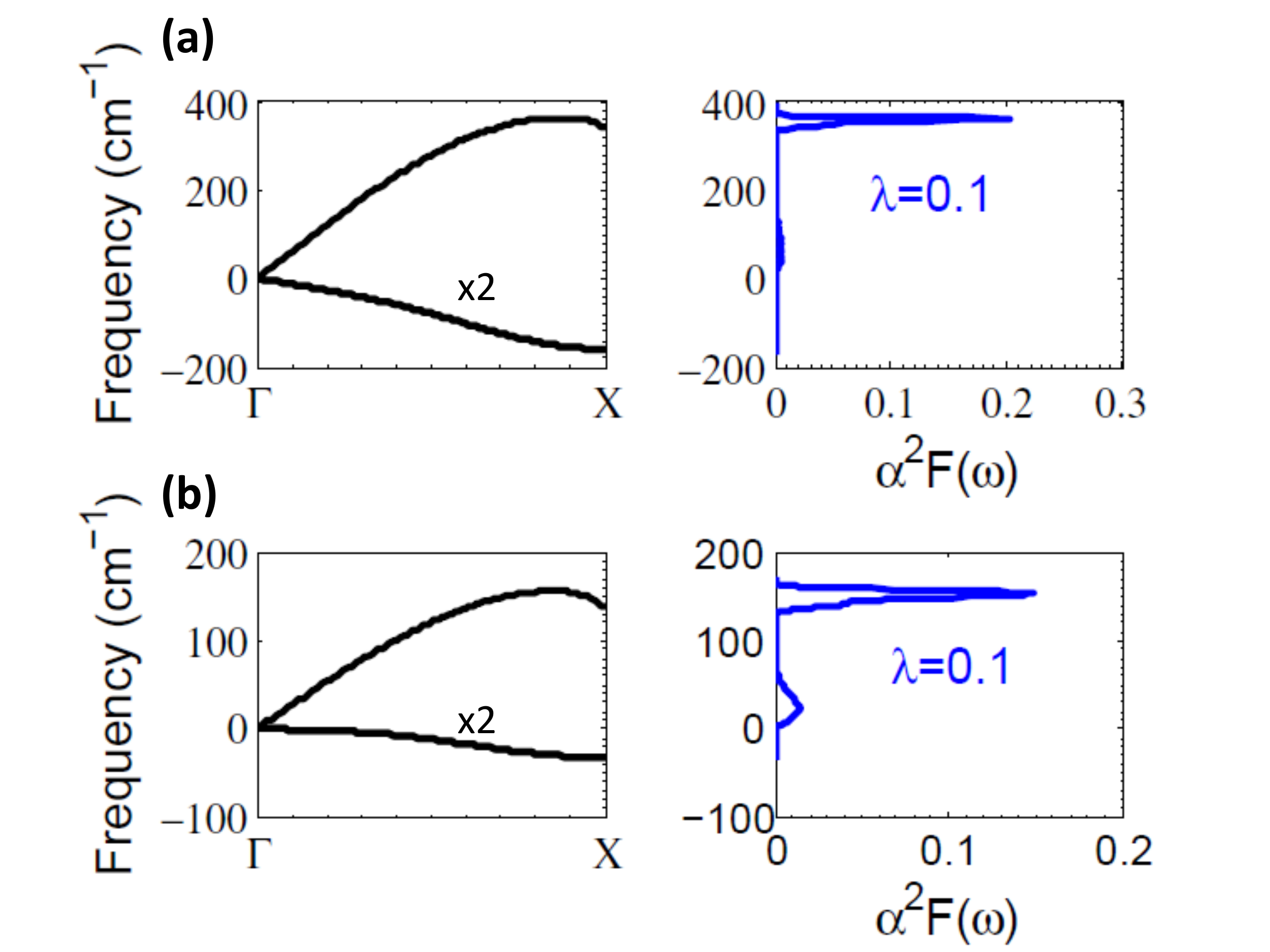}
\end{center}
\caption {The phonon band structure, Eliashberg Spectral Function and electron-phonon coupling constant for atomic wires of (a) lithium and (b) sodium.}
\end{figure}
The phonon band structure shows the presence of one longitudinal acoustic phonon mode and two degenerate transverse acoustic modes, due to the one-dimensional symmetry of the wires. The transverse acoustic phonon modes dip into imaginary frequencies, which indicates a preference for the wires to distort into a zig-zag configuration. Imaginary frequency transverse phonon modes have been observed in atomic wires composed of of Al, Cu, Ag, Cu and Pb\cite{verstraete2006phonon,sen2006peierls,simbeck2012aluminum} and their degeneracy has been labelled accordingly in the figure. The spectral function shows peaks only around the longitudinal phonon modes, indicating that electrons do not couple at all to the imaginary frequency transverse modes. Since the longitudinal phonons have a much larger absolutely value of frequency than the transverse phonons, the factor of $1/\omega$ results in a weak electron-phonon coupling of only $\lambda = 0.1$ for both metals. This number is larger than the coupling constants in Al and Ag atomic wires but smaller than the coupling constants in Cu and Au atomic wires\cite{simbeck2012aluminum}. 

Comparing these numbers to the electron-phonon coupling in the bulk, we see that both metals show a reduced overall coupling in the atomic wire configuration. The bulk coupling for BCC Li is $\lambda=0.45$\cite{liu1991electron,profeta2006superconductivity}while for Na it is $\lambda=0.18$ \cite{bauer1998electron}. The sizeable reduction in electron-phonon coupling is attributed to the strong effects of quantum confinement in one-dimension. 

We note that while results for the electron-phonon coupling have been published\cite{sen2006peierls}, the k-point sampling was no dense enough to attain an accurate value of the electron-phonon coupling constant. We have reproduced the published results using the reported k-point grid ($1\times 1 \times 24$, resulting in a coupling constant of $\lambda=0.01$) but note that our k-point sampling of $1\times 1 \times 128$ gives a more accurate and fully-converged value of the electron-phonon coupling constant $\lambda$ ($\lambda=0.1$). 

It is worth noting that the identical (to one decimal place) values of the electron-phonon coupling constants in the Li and Na atomic wires is coincidental. While the maximum phonon frequencies are largest in the Li wire (extending up to 400 cm$^{-1}$) the peak heights of the spectral function are also larger (rising to around $\alpha^2F(\omega)=0.2$). In the case of the Na atomic wire, the maximum phonon frequencies are smaller (extending only to around 150 cm$^{-1}$) but the spectral function only rises to around $\alpha^2F(\omega)$=0.15, yielding a very similar overall coupling strength. 

\section{Conclusion} 
In conclusion, we have shown that weak electron-phonon coupling prevails in single-atom thick metallic wires composed of the early alkali metals Li and Na. The weak coupling is the result of the s-like electons at the Fermi level coupling exclusively to the longitudinal acoustic phonon branch, which has a larger magnitude of frequency than the transverse modes. The higher frequencies are effectively damped by the factor of $1/\omega$ in the formula for the electron-phonon coupling constant, resulting in very small overall coupling. When compared to earlier results on other metals, a trend becomes evident in which s- and p-like electrons at the Fermi level result in weak electron-phonon coupling in one-dimension, while the presence of d-like electrons results in stronger coupling due to the involvement of the transverse acoustic phonons. Future studies will focus on the calculation of the electron-phonon coupling for atomic wires composed of more complicated metals as well as studies of the electron-phonon coupling at contacts and interfaces. 

This work used computational resources provided by the Computational Center for Nanotechnology Innovations (CCNI) at Rensselaer Polytechnic Institute. The authors acknowledge support by the Army Research Lab Multiscale Multidisciplinary Modeling of Electronic Materials (MSME) Collaborative Research Alliance (CRA) and the INDO-US Forum. 

\bibliography{Alkali}

\begin{thebibliography}{10}
\expandafter\ifx\csname url\endcsname\relax
  \def\url#1{\texttt{#1}}\fi
\expandafter\ifx\csname urlprefix\endcsname\relax\def\urlprefix{URL }\fi

\bibitem{Ashcroft1976solid}
N.~Ashcroft, N.~Mermin, {Solid State Physcs }, Thomson Learning, 1976.

\bibitem{savrasov1994linear}
S.~Savrasov, D.~Savrasov, O.~Andersen, Linear-response calculations of
  electron-phonon interactions, Phys. Rev. Lett. 72~(3) (1994) 372--375.

\bibitem{savrasov1996electron}
S.~Y. Savrasov, D.~Y. Savrasov, Electron-phonon interactions and related
  physical properties of metals from linear-response theory, Phys. Rev. B 54
  (1996) 16487--16501.

\bibitem{bauer1998electron}
R.~Bauer, A.~Schmid, P.~Pavone, D.~Strauch, Electron-phonon coupling in the
  metallic elements al, au, na, and nb: A first-principles study, Phys. Rev. B
  57 (1998) 11276--11282.

\bibitem{Frederiksen2004inelastic}
T.~Frederiksen, M.~Brandbyge, N.~Lorente, A.-P. Jauho, Inelastic scattering and
  local heating in atomic gold wires, Phys. Rev. Lett. 93 (2004) 256601.

\bibitem{Picaud2003phonons}
F.~Picaud, A.~D. Corso, E.~Tosatti, Phonons softening in tip-stretched
  monatomic nanowires, Surface Science 532–535~(0) (2003) 544 -- 548.

\bibitem{delaVega2006universal}
L.~de~la Vega, A.~Martin-Rodero, N.~Agrait, A.~L. Yeyati, Universal features of
  electron-phonon interactions in atomic wires, Phys. Rev. B 73 (2006) 075428.

\bibitem{Agrait2002onset}
N.~Agrait, C.~Untiedt, G.~Rubio-Bollinger, S.~Vieira, Onset of energy
  dissipation in ballistic atomic wires, Phys. Rev. Lett. 88 (2002) 216803.

\bibitem{Agrait2002electron}
N.~Agrait, C.~Untiedt, G.~Rubio-Bollinger, S.~Vieira, Electron transport and
  phonons in atomic wires, Chemical Physics 281~(2–3) (2002) 231 -- 234.

\bibitem{bohler2009point}
T.~Boehler, A.~Edtbauer, E.~Scheer, Point-contact spectroscopy on aluminum
  atomic-size contacts: longitudinal and transverse vibronic excitations, New
  Journal of Physics 11~(013036).

\bibitem{verstraete2006phonon}
M.~J. Verstraete, X.~Gonze, Phonon band structure and electron-phonon
  interactions in metallic nanowires, Phys. Rev. B 74 (2006) 153408.

\bibitem{sen2006peierls}
P.~Sen, Peierls instability and electron-phonon coupling in a one-dimensional
  sodium wire, Chemical Physics Letters 428~(4-6) (2006) 430--435.

\bibitem{simbeck2012aluminum}
A.~J. Simbeck, N.~Lanzillo, N.~Kharche, M.~J. Verstraete, S.~K. Nayak, Aluminum
  conducts better than copper at the atomic scale: A first-principles study of
  metallic atomic wires, ACS Nano 6~(12) (2012) 10449--10455.

\bibitem{kohn1964inhomogeneous}
P.~Hohenberg, W.~Kohn, Inhomogeneous electron gas, Phsyical Review 136.

\bibitem{kohn1965self}
W.~Kohn, L.~Sham, Self-consistent equations including exchange and correlation
  effects, Physical Review 140~(4A) (1965) 1133--\&.

\bibitem{gonze1997dynamical}
X.~Gonze, C.~Lee, Dynamical matrices, born effective charges, dielectric
  permittivity tensors, and interatomic force constants from density-functional
  perturbation theory, Physical Review B 55~(16) (1997) 10355--10368.

\bibitem{gonze2002first}
X.~Gonze, J.~Beuken, R.~Caracas, F.~Detraux, M.~Fuchs, G.~Rignanese, L.~Sindic,
  M.~Verstraete, G.~Zerah, F.~Jollet, M.~Torrent, A.~Roy, M.~Mikami, P.~Ghosez,
  J.~Raty, D.~Allan, First-principles computation of material properties: the
  abinit software project, Computational Materials Science 25~(3) (2002)
  478--492.

\bibitem{gonze2005brief}
X.~Gonze, G.~Rignanese, M.~Verstraete, J.~Beuken, Y.~Pouillon, R.~Caracas,
  F.~Jollet, M.~Torrent, G.~Zerah, M.~Mikami, P.~Ghosez, M.~Veithen, J.~Raty,
  V.~Olevano, F.~Bruneval, L.~Reining, R.~Godby, G.~Onida, D.~Hamann, D.~Allan,
  A brief introduction to the abinit software package, Zeitschrift Fur
  Kristallographie 220~(5-6) (2005) 558--562.

\bibitem{gonze2009abinit}
X.~Gonze, B.~Amadon, P.~M. Anglade, J.~M. Beuken, F.~Bottin, P.~Boulanger,
  F.~Bruneval, D.~Caliste, R.~Caracas, M.~Cote, T.~Deutsch, L.~Genovese,
  P.~Ghosez, M.~Giantomassi, S.~Goedecker, D.~R. Hamann, P.~Hermet, F.~Jollet,
  G.~Jomard, S.~Leroux, M.~Mancini, S.~Mazevet, M.~J.~T. Oliveira, G.~Onida,
  Y.~Pouillon, T.~Rangel, G.~M. Rignanese, D.~Sangalli, R.~Shaltaf, M.~Torrent,
  M.~J. Verstraete, G.~Zerah, J.~W. Zwanziger, Abinit: First-principles
  approach to material and nanosystem properties, Computer Physics
  Communications 180~(12) (2009) 2582--2615.

\bibitem{troullier1991efficient}
N.~Troullier, J.~Martins, Efficient pseudopotentials for plane-wave
  calculations .2. operators for fast iterative diagonalization, Physical
  Review B 43~(11) (1991) 8861--8869.

\bibitem{liu1991electron}
A.~Liu, M.~Cohen, Electron-phonon coupling in bcc and 9r lithium, Phys. Rev. B
  44 (1991) 9678.

\bibitem{profeta2006superconductivity}
G.~Profeta, C.~Franchini, N.~Lathiotakis, A.~Floris, A.~Sanna, M.~Marques,
  M.~Luders, S.~Massidda, E.~Gross, A.~Continenza, Superconductivity in
  lithium, potassium and aluminum under extreme pressure: A first-principles
  study, Physical Review Letters 96 (2006) 047003.

\end{thebibliography}

\end{document}